\documentclass[twoside,twocolumn,9pt]{article}
\usepackage{extsizes}
\usepackage[super,sort&compress,comma]{natbib} 

\usepackage[version=3]{mhchem}
\usepackage[left=1.5cm, right=1.5cm, top=1.785cm, bottom=2.0cm]{geometry}
\usepackage{balance}
\usepackage{times,mathptmx}
\usepackage{sectsty}
\usepackage{graphicx} 
\usepackage{lastpage}
\usepackage[format=plain,justification=justified,singlelinecheck=false,font={stretch=1.125,small,sf},labelfont=bf,labelsep=space]{caption}
\usepackage{float}
\usepackage{fancyhdr}
\usepackage{fnpos}
\usepackage[english]{babel}
\usepackage{array}
\usepackage{droidsans}
\usepackage{charter}
\usepackage[T1]{fontenc}
\usepackage[usenames,dvipsnames]{xcolor}
\usepackage{setspace}
\usepackage[compact]{titlesec}
\usepackage{gensymb}
\usepackage{cuted}
\usepackage{tgbonum}
\usepackage{wrapfig}
\usepackage[utf8x]{inputenc}

\usepackage{epstopdf}
\usepackage{gensymb}

\definecolor{cream}{RGB}{222,217,201}

\begin{document}
\begin{strip}
\textbf{\Huge{Juggling bubbles in square capillaries: an experimental proof of non-pairwise bubble interactions}}\\
\newline
\noindent\Large{Ga\"el Ginot,\textit{$^{a}$} Reinhard H\"ohler\textit{$^{b, c}$}, Sandrine Mariot \textit{$^{d}$}, Andy Kraynik \textit{$^{a,d}$} and Wiebke Drenckhan \textit{$^{a,d}$}} 
\newline\newline
 \raggedright
\footnotesize{\textit{$^{a}$~Institut Charles Sadron, CNRS UPR22 - Universit\'{e} de Strasbourg, Strasbourg, France;  Fax: 33 (0)3 88 41 40 99; Tel: 33 (0)3 88 41 40 43; E-mail: Wiebke.Drenckhan@ics-cnrs.unistra.fr}\\
\textit{$^{b}$~Sorbonne Universit\'{e}s, UPMC Univ Paris 06, CNRS-UMR 7588, Institut des NanoSciences de Paris, 4 place Jussieu, 75005 Paris, France}\\
\textit{$^{c}$~Universit\'{e} Paris-Est Marne-la-Vall\'{e}e , 5 Bd Descartes, Champs-sur-Marne, F-77454 Marne-la-Vall\'{e} cedex 2, France. }\\
\textit{$^{d}$~Laboratoire de Physique des Solides, CNRS, Univ. Paris-Sud, Universit\'{e} Paris-Saclay, 91405 Orsay Cedex, France. }}
\newline \newline \noindent \Large{\textbf{Abstract.}}\large{The physical properties of an ensemble of tightly packed particles like bubbles, drops or solid grains are controlled by their interactions. For the case of bubbles and drops it has recently been shown theoretically and computationally that their interactions cannot generally be represented by pair-wise additive potentials, as is commonly done for  simulations of soft grain packings. This has important consequences for the mechanical properties of foams and emulsions, especially for strongly deformed bubbles or droplets well above the jamming point. Here we provide the first experimental confirmation of this prediction by quantifying the interactions between bubbles in simple model foams consisting of trains of equal-volume bubbles confined in square capillaries. The obtained interaction laws agree quantitatively with Surface Evolver simulations and are well described by an analytically derived expression based on the recently developed non-pairwise interaction model of H\"ohler \textit{et al.} [{Soft Matter}, 2017, \textbf{13},(7):1371], based on Morse-Witten theory. While all experiments are done at Bond numbers sufficiently low for the hydrostatic pressure variation across one bubble to be negligible, we provide the full analysis taking into account gravity in the appendix for the interested reader. Even though the article focuses on foams, all results directly apply to the case of emulsions.} 
\end{strip}
\pagestyle{fancy}
\thispagestyle{plain}
\makeatletter 
\newlength{\figrulesep} 
\setlength{\figrulesep}{0.5\textfloatsep} 
\newcommand{\topfigrule}{\vspace*{-1pt}%
\noindent{\color{cream}\rule[-\figrulesep]{\columnwidth}{1.5pt}} }
\newcommand{\botfigrule}{\vspace*{-2pt}%
\noindent{\color{cream}\rule[\figrulesep]{\columnwidth}{1.5pt}} }
\newcommand{\dblfigrule}{\vspace*{-1pt}%
\noindent{\color{cream}\rule[-\figrulesep]{\textwidth}{1.5pt}} }
\makeatother

\section{Introduction}

Foams consist of bubbles which are tightly packed within a liquid \cite{cantat2013foams,weaire2001}. Their properties depend strongly on the volume fraction $\phi$ of the continuous phase. For $\phi$ above a critical value (typically $\phi_c\approx 0.36$), the dispersion behaves as a liquid and the bubbles tend to have a spherical shape that minimises their interfacial energy. At jamming ($\phi$ = $\phi_c$), the bubbles are still spherical yet sufficiently in contact with each other so that the system behaves as a soft solid: even in the presence of stress, a static mechanical equilibrium is reached \cite{VanHecke2010,Liu2010}.  Taking into account the absence of static friction between bubbles, constraint counting arguments show that in such a packing bubbles touch on average $\left< z \right> \approx 6$ neighbours \cite{VanHecke2010,Song2007}. As liquid is removed from such a "wet" foam, bubbles deform against each other, creating  more and more contacts through which the interaction forces  are transmitted. Their magnitude is given by the product of the contact area and the Laplace overpressure in the gas\cite{cantat2013foams,weaire2001}, set by the interfacial curvature and the surface tension $\gamma$ . These interaction forces control how exactly bubbles pack and move against each other, having an important influence on the elastic, yielding, and plastic flow properties of foams.

Inspired by work on granular media, the interaction forces between bubbles slightly beyond jamming are often assumed to be pair-wise additive and they are modeled by power laws. The deformation of the contact between two neighboring bubbles with center positions $\vec{r}_1, \vec{r}_2$ and undeformed radii $R_1,R_2$ is described by a deformation parameter
\begin{equation}
\label{eq:overlap}
x= \frac{\left| \vec{r}_1-\vec{r}_2\right|}{(R_1+R_2)}-1,
\end{equation}
and the force is assumed to vary with $x$ following a power law
\begin{equation}
f \sim x^{\alpha-1}.
\end{equation}
For example, $\alpha = 2$ if the contact behaves like a repulsive harmonic spring or $\alpha = \frac{5}{2}$ for a Hertzian interaction\cite{VanHecke2010,Lacasse1996,Seth2010}.

However, unlike solid grains, bubbles are easily deformed and their volume remains constant upon compression by their neighbours. This is so because capillary forces are too small to measurably compress gas under atmospheric conditions. Hence, the deformation of a bubble at one contact impacts the deformation on any other point of the bubble.

The volume conservation thus leads to "non-pairwise" (or "many-body") interactions between bubbles: there is no unique relation between displacement and force at a given contact. Analytical calculations and Surface Evolver simulations have shown that this has a significant impact on mechanical  properties of monodisperse and weakly polydisperse foams  \cite{Hohler2017}. Similar non-pairwise interactions have been reported for 2D granular materials consisting of soft elastic disks \cite{Siber2003}.\par

Non-pairwise interactions between two objects do not only depend on their relative position, but also on the positions of other neighbouring objects. Such interactions have already been reported in other fields of soft matter physics. An example are  charged colloidal particles in a salt solution. They perturb the ionic concentrations in their neighbourhood so that their charges are screened off. Experiments have demonstrated that the interaction between two such particles is perturbed if a  third particle is approached which modifies the ionic distributions \cite{Merrill2009}. To deal with this complexity in models, many-body interactions are often represented in terms of an "effective two-body interaction", expected to be valid on average. As we will demonstrate, such an approach can fail for foams and emulsions.\par

Based on a seminal paper by Morse and Witten\cite{Morse1993}, H\"ohler {\it et al.} \cite{Hohler2017} have established force/displacement relationships for multiple bubble contacts in a slightly polydisperse, 3D foam with periodic boundary conditions. They show and explain significant deviation from pair-wise interaction behaviour, combining theoretical calculations and Surface Evolver simulations. This was extended to polydisperse, disordered 2D foams by Weaire {\it et al.} \cite{Weaire2017}. Only very close to the jamming point where bubbles barely touch, volume conservation constraints become negligible and, asymptotically, pair-wise ("two body") interaction behavior is reached. In this limit, in 3D systems, a logarithmic softening of the interaction law has been evidenced experimentally \cite{aussillous2006,Chevy2012} and justified theoretically\cite{Morse1993,Chevy2012,Chan2011} 
\begin{equation}
\label{eq:MWasym}
x \sim f\, \ln f.
\end{equation}
We see that the effective stiffness of the contact, given by $f/x$, goes to zero as $1/\ln f$ in the limit of small forces. The fundamental mechanism leading to these features was first pointed out by Morse and Witten \cite{Morse1993}, yet insight into its full implication for jamming behavior as well as foam and emulsion science is only emerging now.

\begin{figure}[h]
\centering
  \includegraphics[width=8.5cm]{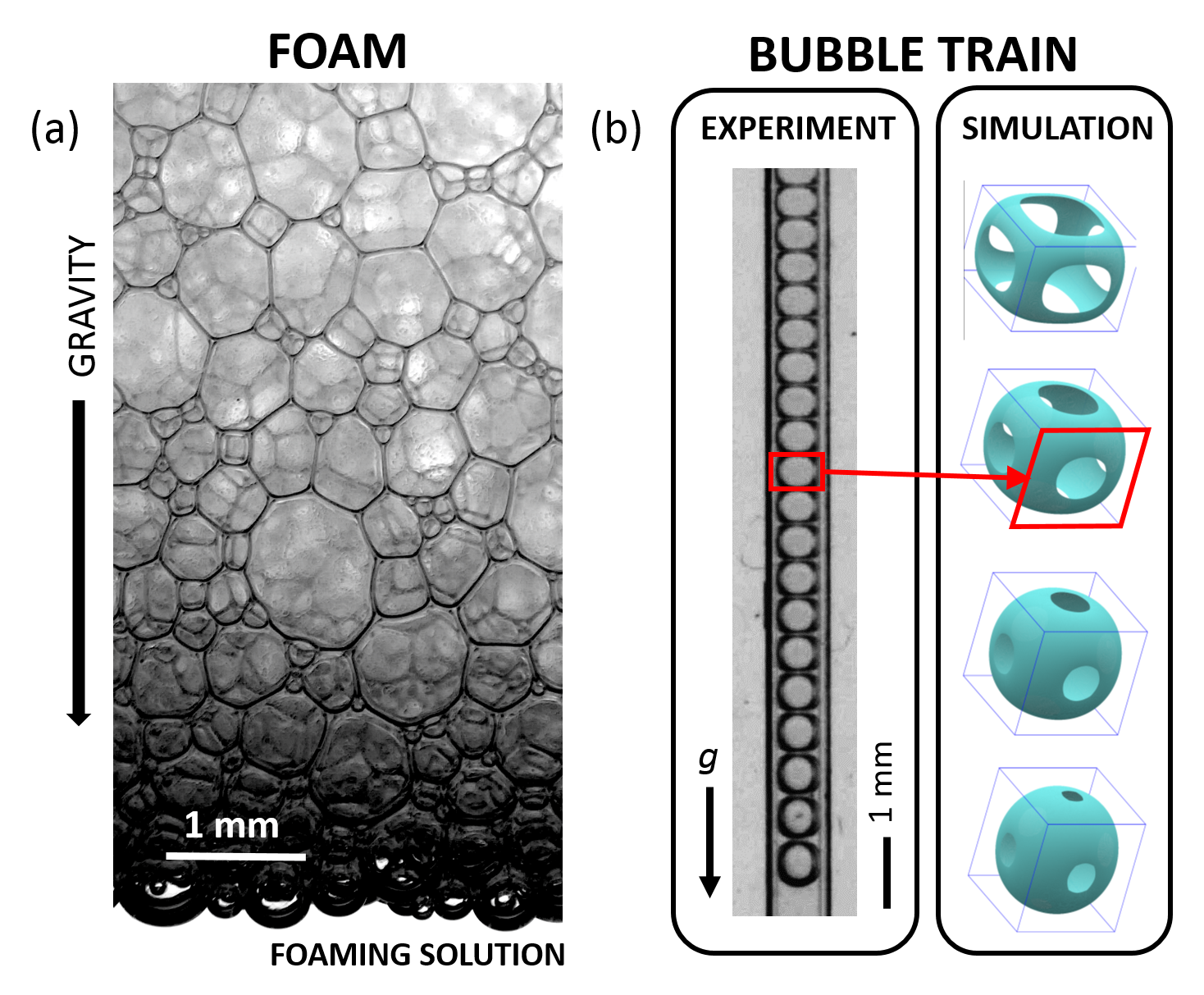}
  \caption{(a) Photograph of a typical foam floating on a foaming solution, showing the progressive deformation of the bubbles with increasing foam height. (b) Simplified model foam (left: experiment; right: simulation) studied in this article consisting of a train of equal-volume bubbles confined in a capillary with square cross-section.} 
  \label{fig:Foamfoto}
\end{figure}

While theoretical and computational investigations into the matter have advanced and converged significantly over the last years \cite{Weaire2017, Hohler2017, Hohler2018}, experimental verifications so far exist only for the case of individual droplets in contact with solid surfaces \cite{aussillous2006,Chevy2012,Hutzler2018}. The goal of this paper is therefore to establish the first experimental confirmation and quantification of the many-body nature of bubble interactions. Since these measurements require accurate control of the bubble contacts, we work with a simplified foam structure, which consists of a train of equal-volume bubbles confined in a capillary with square cross-section, as shown in Fig. \ref{fig:Foamfoto}b. Each bubble has therefore six contacts, four with the lateral walls, and two with its neighbours. We can vary the confinement due to the wall contacts by changing the ratio between the bubble and the capillary size. If the capillary is held vertically, each bubble is pushed upwards by buoyancy. The force at a given contact between neighboring bubbles is thus set by the cumulative buoyancy of all bubbles below. These forces can be modulated by tilting the capillary. If the bubble size is much smaller than the so called capillary length  $\sqrt{\gamma/\Delta\rho g}$, the deformation of individual bubbles is  dominated by the contact forces while the impact on the pressure gradient in the liquid on the bubble shape is negligible. $\Delta\rho$ is the difference between the densities of the internal and continuous phases, $g$ the acceleration due to gravity. In terms of the Bond number,  defined as $Bo = \frac{\Delta\rho g R_0^2}{\gamma }$, the same condition can be expressed as $ Bo \ll 1$.

To investigate the mechanical response of the bubbles along the bubble train, we observe for each of them the distance between the two contacts with their neighbors (Section \ref{sec:Methods}). The bubble-bubble contact force due to buoyancy can  be deduced from the observed gas volume below a given bubble in the train and this force can be tuned by tilting the capillary.  We are thus able to probe the many-body character of bubble interactions experimentally. Our experimental results (Section \ref{sec:results}) confirm quantitatively the theoretical and computational predictions derived in Section \ref{sec:TheoSim}, both showing significant deviations from a pair-wise additive power-law behavior (Section \ref{sec:twobody}). While the interpretation of our experimental results neglects the gravity-pressure variation over the scale of one bubble, we provide the full gravitational model in APPENDIX \ref{sec:theorygravity} for the interested reader. \par

Even though our experiments were carried out in a confined quasi 1D geometry, the interaction law that we validate can be used to describe bubble interactions in 2D systems such as microfluidic channels or in 3D foams. Our results highlight the need to take into account explicitly the non-pairwise coupling between contacts.\par 

While our investigations were performed on bubbles, all results can be directly translated to emulsions. 


\label{sec:TheoSim}
\begin{figure}[h!]
\centering
  \includegraphics[width=7.5cm]{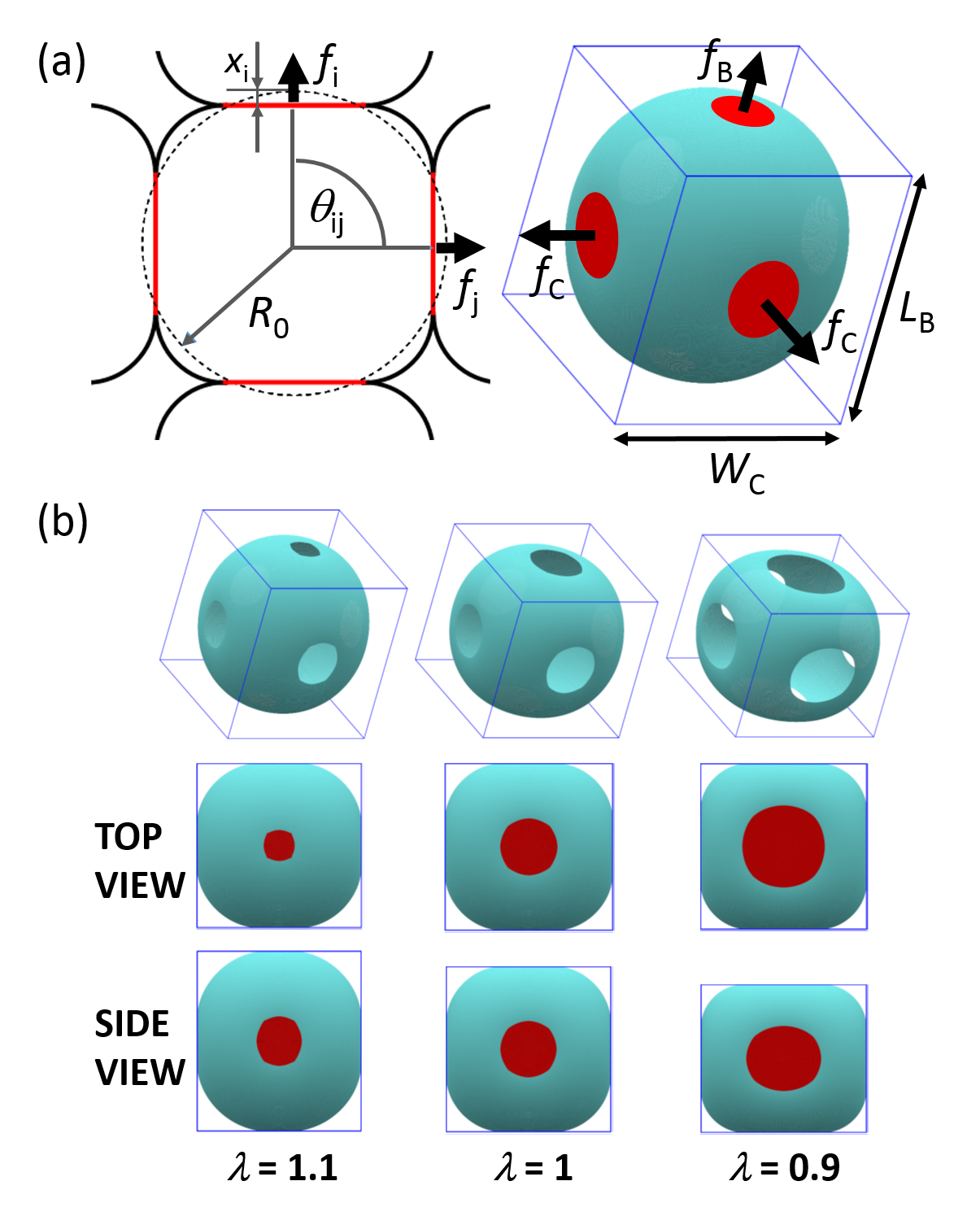}
  \caption{(a) Left: General notation used in this article for the description of the bubble geometry and the associated contact forces $f_i$ which are exerted radially with respect to the bubble centre. Right: Repeat unit of bubbles confined in a capillary, outlined by a frame. $f_C $ and $f_B $ are, respectively, the forces radially exerted by the bubble on the walls of the capillary and on its two neighbors. $W_c $ is the distance between opposite walls and $L_B$ the length of the confined bubble. (b) Views of Surface Evolver simulations (Section \ref{sec:evolver}) for three bubble aspect ratios $\lambda = L_B/W_c $ (cf. (a)), for a confinement ratio $W_c/R_O= 1.81$.  The second and third line shows top and side views where the contacts of the bubble are highlighted in red.}
  \label{fgr:Variables}
\end{figure}

\section{Modelling}
\subsection{Effective two-body approach \label{sec:twobody}}

Even though it is well known that bubble interactions are non-pairwise due to the conservation of bubble volume \cite{Morse1993, Mason1997,Buzza1994}, the difficulties of implementing accurate theories of this effect have motivated the search for an effective two-body interaction law for bubbles. Lacasse {\textit{et al.}} performed Surface Evolver simulations (Section \ref{sec:evolver}) for bubbles subjected {\it only to  isotropic} compression, and deduced from these data the repulsive interaction potentials in the form of a power law \cite{Mason1997, Lacasse1996}. In terms of normalised contact forces $f(x)$, this law takes the following form where $z$ is the bubble coordination number
\begin{alignat}{2}
\label{eq:masonlaw}
f(x)= \left\{ 
\begin{array}{l l}
\kappa(z ) \alpha(z ) \frac{\left( (1+x)^{-3}-1\right)^{\alpha(z ) - 1} }{ (1+x)^{4}} & \quad \text{for $x <0$}\\
\\
0 & \quad \text{for $x \ge 0$.} \\ \end{array} \right. 
\end{alignat}
The functions $\kappa(z )$ and $\alpha(z )$ are fitted to optimise agreement with Surface Evolver simulation data \cite{Mason1997, Lacasse1996}   \footnote{In Table 1 of the pioneering publication by Lacasse {\it et al.} \cite{Lacasse1996}   $\kappa$  was expressed in terms of a constant $C=\kappa /( 12\pi)$}. The interaction is predicted to be approximately harmonic ($\alpha = 2$) for $z = 6$ and increasingly anharmonic for larger coordination numbers.     For 6 contacts, the case relevant for our experiments as well as random foams at the jamming transition, the predicted values of the parameters in Eq. (\ref{eq:masonlaw}) are $\kappa=0.78$ and $\alpha=2.2$.  In the effective two-body approach, the same equation is applied independently to all contacts, ignoring non-pairwise effects. The experiments reported here provide a test for this approximation, in the case of 1D bubble trains.

\subsection{Many-body approach: Morse-Witten theory \label{sec:MW}}

We focus on static foams with approximately spherical bubbles and purely repulsive interactions. Polydispersity is assumed to be very weak so that the bubble-bubble contacts are to a good approximation flat \cite{Hohler2018}. All contact forces are assumed to be radial. Bubble contacts with a fully wetted wall or with a neighboring bubble are described in the same framework. They are labeled by an index $i$, as sketched in Fig. \ref{fgr:Variables}. We choose a coordinate system whose center coincides with the bubble's centroid and define an effective radius $R_0$ as the radius of a sphere whose volume is the same as that of the bubble.  All forces are divided by $\gamma R_0$ to make them dimensionless. The surface tension $\gamma$ is assumed to be constant with deformation, as is the case for the low-molecular weight surfactants used in the accompanying experiments (Section \ref{sec:Methods}). Depending on the contact forces, the distance of each contact from the bubble's centroid will deviate from $R_0$. The associated dimensionless deformations  $x_i$ of the contact zones are  given by Eq. (\ref{eq:overlap}) (with $R_1 = R_2 = R_0$) for bubble-bubble contacts.
  A calculation  based on the linearized Laplace equation yields the following interaction law, relating contact forces $f_i$ and contact deformations $x_i$\cite{Morse1993,Hohler2017,Hohler2018} 
\begin{equation}
x_i  =   \frac{1}{24\pi} \left[5 +6\ln\left(\frac{f_i}{8\pi}\right) \right]  f_i -  \sum\limits_{j\neq i}G(\theta_{ij})  f_j.
 \label{eq:interactionlaw}
\end{equation} 
$\theta_{ij}$ is the angle between vectors pointing from the centroid of the bubble towards the centers of the contacts $i$ and $j$. The Green's function 
\begin{equation}
 \label{eq:green}
G(\theta)=-\frac{1}{4\pi} \left\{ \frac{1}{2}+\frac{4}{3}\cos\theta+\cos\theta \ln\left[\sin^2(\theta/2)\right] \right\}
\end{equation}
 describes the shape changes induced by a  force locally acting on a bubble.
Due to the linearisation of the Laplace equation used in the derivation, Eqs. (\ref{eq:interactionlaw}) and (\ref{eq:green})  hold only for sufficiently small dimensionless forces  $f_i < 1$, i.e. for weakly deformed bubbles.
The first term in Eq. (\ref{eq:interactionlaw}) represents the deformation $x_i$ due to the force $f_i$ acting at the same contact. We see that this local contribution corresponds to Eq. (\ref{eq:MWasym}) in the limit of small forces. The second term in Eq. (\ref{eq:interactionlaw}) gives the non-pairwise contribution to the deformation $x_i$, resulting from the forces $f_j$ at the contacts $j \neq i $. In the deeply jammed regime of foams or emulsions ($\phi << \phi_c$), this second term gives rise to many-body interaction behaviour whereas very close to the jamming point where forces go to zero ($\phi \approx \phi_c$), the first term dominates.

In Morse-Witten theory, Eq. (\ref{eq:green}) is derived analytically from the shape of a bubble buoyed against a flat horizontal plate, due to a pressure gradient induced by gravity\cite{Morse1993}. On this basis, the response to multiple contact forces in different directions given in Eq. (\ref{eq:interactionlaw}) is constructed using the principle of superposition: the deformations of the interface due to the different forces are added. In the case where the contact force vectors add up to zero, the pressure gradients associated with each force cancel. This results in a model for a bubble  deformed only by contact forces in the absence of pressure gradients (gravity). 
\par Without any modification Eqs. (\ref{eq:interactionlaw}) and (\ref{eq:green})  also apply if in addition to contact forces, the bubble is deformed by a gravity-induced pressure gradient, as  pointed out in \cite{Hohler2018}.
In this case forces specified in Eq. (\ref{eq:interactionlaw}) do not add up to zero and the buoyancy-induced pressure gradient provides the missing force that ensures the force balance. We obtain in this case the solution for a bubble subjected to a combination of contact forces and pressure gradients. This is presented in more detail in the Appendix \ref{sec:theorygravity}.\par
We now apply Morse-Witten theory to our experimental configuration: a train of equal-volume bubbles in static equilibrium, confined in a capillary of square cross-section with side length $W_c$, as shown in the photograph of Fig. \ref{fig:Foamfoto}b and in the schematic of Fig. \ref{fgr:Variables}a.  Each bubble is confined by two neighbouring bubbles (contact forces $f_B$) and by four capillary walls (contact forces $f_C$). This makes in total six confining contacts per bubble. We assume zero contact angle for all contacts (perfect wetting). 
\par In the limit of small Bond numbers $Bo \rightarrow 0$ all bubble-wall contacts are equivalent by symmetry and the same holds for the two bubble-bubble contacts. This makes the predictions of Morse-Witten theory  relatively simple to calculate.\par 
The length of the bubble along the axis of the tube is called $L_B$. In dimensionless form, this length and the capillary width $W_C$ are expressed as an axial length $L_B/ 2 R_0$ and a bubble confinement ratio $W_C/2R_0$. These two experimental parameters determine the contact deformations $x_i$ used in the theory. We distinguish the deformation $x_B$ at bubble-bubble contacts and $x_C$ at bubble-wall 
\begin{align}
\label{eq:XB}
x_B=\frac{L_B}{2R_o}-1 \\
\label{eq:XC}
x_C=\frac{W_C}{2R_o}-1.
\end{align}
For a spherical bubble that exactly fits in the capillary $(W_C/2R_0 =1)$ and in the space between its top and bottom neighbours $(L_B/2R_0=1)$, the deformations $x_C$ and $x_B$ are thus by definition zero. If the space available in the lateral or axial directions of the capillary is smaller than $2R_o$, the respective contact strains become increasingly negative. The anisotropy of the bubble deformation may be expressed by the aspect ratio $\lambda $ defined as 
\begin{equation}
\label{eq:deflambda}
\lambda= \frac{L_B}{W_C}=\frac{1+x_B}{1+x_C}.
\end{equation}
Fig. \ref{fgr:Variables} illustrates the shapes of bubbles for different aspect ratios $\lambda$ for a confinement ratio $W_C/2R_0 = 0.91$ obtained from Surface Evolver simulations (Section \ref{sec:evolver}).

Under the assumptions stated above, the general interaction law of Eq. (\ref{eq:interactionlaw})  yields the following relations
\begin{align}
\label{eq:XBF}
x_B&=  \frac{1}{24\pi} \left[5 +6\ln\left(\frac{f_B}{8\pi}\right) \right]  f_B -  4 G(\pi/2)  f_C-   G(\pi)  f_B\\
\label{eq:XCf}
x_C&= \frac{1}{24\pi} \left[5 +6\ln\left(\frac{f_C}{8\pi}\right) \right]  f_C -   2 G(\pi/2)  f_C -   G(\pi)  f_C -2 G(\pi/2)  f_B.
\end{align}
Using Eq. (\ref{eq:green}) we find
\begin{align}
\label{eq:Gpi2}
 G(\pi/2)=-\frac{1}{8 \pi}
\\
\label{eq:Gpi}
 G(\pi)=\frac{5}{24 \pi}
\end{align}
and we simplify Eqs. (\ref{eq:XBF})-(\ref{eq:XCf}) 
\begin{align}
\label{eq:XBF2}
x_B &=  \frac{1}{4\pi} \ln\left(\frac{f_B}{8\pi}\right)   f_B +\frac{1}{2 \pi}  f_C\\
\label{eq:XCf2}
x_C &= \frac{1}{4\pi} \ln\left(\frac{f_C}{8\pi} e\right)  f_C +  \frac{1}{4 \pi}  f_B.
\end{align}
To express $f_C$ as a function of the given quantities $f_B$ and $x_C$ Eq. (\ref{eq:XCf2}) is solved using the Lambert function $W$ \cite{Corless1996}. Among its different branches, the Lambert function of index $-1$ has the physically correct asymptotic behavior given in Eq. (\ref{eq:MWasym}). We therefore obtain
\begin{equation}
\label{eq:XCf3}
f_C = \frac{4\pi x_c-f_B}{W_{-1}\left(\frac{(4\pi x_C-f_B)e}{8\pi}\right)}. \\
\end{equation}
Inserting Eq. (\ref{eq:XCf3})  into Eq. (\ref{eq:XBF2}) yields an expression for $x_B$ 
\begin{equation}
\label{eq:XBF3}
x_B =  \frac{1}{4\pi} \ln\left(\frac{f_B}{8\pi}\right)   f_B +\frac{1}{ 2\pi} \frac{4\pi x_c-f_B}{W_{-1}\left(\frac{(4\pi x_c-f_B)e}{8\pi}\right)}.
\end{equation}

The constant $e$ is the exponential function of 1. The Lambert function is implemented in most mathematical solvers. Its behavior for $x$ close to zero is approximately logarithmic \cite{Corless1996}
\begin{equation}
W_{-1}\left(-x\right) \approx \ln(x) -\ln(-\ln x ).
\end{equation}
However, here we will not use this approximation and proceed instead with an exact solution. 
Expressed in terms of the main experimental parameters ($L_B$,$W_C$ and $R_0$), our result is then
\begin{equation}
\label{eq:XBF4}
\frac{L_B}{2 R_0} =  1 +\frac{1}{4\pi} \ln\left(\frac{f_B}{8\pi}\right)   f_B +\frac{1}{ 2\pi} \frac{4\pi \frac{W_C}{2R_o}-4\pi-f_B}{ W_{-1}\left( \frac{e}{2}(\frac{W_C}{2R_o}-1)-\frac{e}{8\pi}f_B\right)},
\end{equation}
with an expected range of validity\cite{Hohler2018}  $f_B < 1$  .
In the limit where bubble-bubble contact forces go to zero ($f_B \rightarrow 0$) Eq. (\ref{eq:XBF4}) reduces to 
\begin{equation}
\label{eq:aspectratio}
\frac{L_B}{2 R_0} =  1+2\frac{ \frac{W_C}{2R_o}-1}{ W_{-1}\left( \frac{e}{2}(\frac{W_C}{2R_o}-1)\right)}.
\end{equation}

\begin{equation}
\label{eq:deformatop,}
x_B =  \frac{ 2x_C}{ W_{-1}\left(x_C \frac{ e}{2}\right)}.
\end{equation}
We will later compare this result with simulations (Section \ref{sec:evolver}) and experiments (Section \ref{sec:results}) and we will use it to measure the bubble volumes within the capillary (Section \ref{sec:Methods}).

\subsection{Surface Evolver simulations\label{sec:evolver}}

In static equilibrium the interfacial energy of a bubble is minimal  with respect to small variations of its shape, for a fixed bubble volume and a given confinement by walls or neighbors. This principle is the basis of our simulations,  performed using the Surface Evolver\cite{Brakke1992} software, where the gas-liquid interface is represented as an assembly of  finite elements.  \par

\begin{figure}[ht!]
\centering
 \includegraphics[width=8cm]{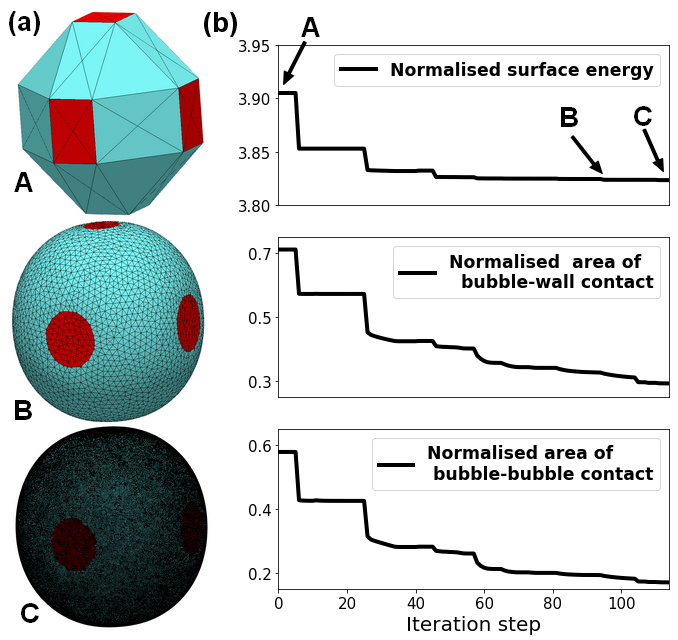}
  \caption{(a) Bubble shapes at different stages during the relaxation of the bubble shape in the Surface Evolver simulation. A - initial geometry with 80 facets. B - 6722 facets. C - final shape with 329.176 facets. (b) Progressive reduction of the energy of the total surface (top) the bubble-wall contact surface (middle) and the bubble-bubble contacts (bottom), upon the iterations. The mesh is considered to be optimal when the relative change in energy is less than $10^{-8}$ between two consecutive iterations of the conjugate gradient procedure.  } 
  \label{fgr:SE_Relax}
\end{figure}

The bubble we simulate may be considered as the repeat unit (unit cell) of a periodic bubble train. We impose the confinement by the walls and the distance between the two bubble-bubble contacts. The contact forces are then determined by multiplying the equilibrium contact areas by the capillary pressure. The pressure in the continuous phase is independent of position in these simulations.\par 
As shown in Fig. \ref{fgr:SE_Relax}a.A, the bubble is initially represented using a very coarse triangular facet meshing so that it appears as a truncated octahedron. Facets confined either by the capillary walls or by the top and bottom neighbor in the capillary are highlighted in red.  
A conjugate gradient algorithm is used iteratively to move the vertices of each facet, so that the total interfacial energy is minimized, respecting the bubble volume and confinement constraints. Upon these iterations, some facet edge lengths and facet areas can become much smaller than average, and such anomalous facets can stall the convergence process. To prevent this from happening, sequences of energy minimisation steps are alternated with the removal of anomalous facets, using mesh optimization tools provided in the Surface Evolver software.\par
When this iterative procedure no longer reduces the total energy, the meshing as a whole is refined by subdividing each facet into smaller triangles. This allows a better representation of curved surfaces, leading to a step-like decrease of the energy. Several examples of this can be been on Fig. \ref{fgr:SE_Relax}.\par
The mesh near the bubble-bubble and bubble-wall contacts is particularly critical. These contacts are rectangular in the initial coarse mesh and must become approximately circular in the fully refined and converged structure. This implies stretching and compression of the mesh near the contact line which generate the anomalous facets mentioned above.  An additional challenge is due to the contact angle which is zero in our system.
This means that in the direction perpendicular to the perimeter of the contacts, the tangent to the gas-liquid interface must become parallel to the wall as the contact line is approached. This curvature of the interface is intrinsically difficult to represent using {\it linear} finite elements. The Surface Evolver also offers quadratic finite elements, but they slow down the computation considerably and using them does not make it straightforward to model the vicinity of the contact line accurately.\par
The curves representing energy versus the number of iteration steps shown in Figure 
\ref{fgr:SE_Relax}b need to be analyzed with great care to judge whether the simulation has converged to the best possible structure for a given mesh refinement, or whether the convergence is stalled by a distorted mesh. The plot at the top of Figure \ref{fgr:SE_Relax}b, showing the total energy of the bubble, suggests that following a global mesh refinement, there is a step-like energy decrease, followed by hardly any significant improvement. Despite this evidence, the iterations must be continued, as illustrated by the two other plots of Figure \ref{fgr:SE_Relax}b focused on the areas of the bubble contacts. They reveal that these parts of the interface continue to evolve slowly but significantly, even after many additional iteration steps. This is due to the difficult convergence of the mesh in the vicinity of the contact line mentioned above. To summarize, the stop criterion for the iterations must be based on convergence of the contact areas, rather then the total energy where the contact contributions can be obscured by the averaging.\par
For  contact angles $\theta > 0$ the gas-liquid interfaces near the contact line is to leading order flat, so that a representation with linear finite elements is adequate. The convergence of the Surface Evolver simulation is in this case much easier to achieve than for $\theta = 0$. We have therefore performed simulations where the surface tension of the gas liquid interface within the contacts were reduced by $10 \%$, $2.5r \%$, and $0 \%$ compared to the value outside the contacts. As shown by Princen, such a decrease of the interfacial energy in the contact region is accompanied by an increase of the contact angle \cite{Princen1965}. For each angle, $\theta > 0$ we determine the contact areas, and extrapolate these results to the case $\theta = 0$ which is the objective of the simulation.    
\par
\begin{figure}[ht!]
\centering
\includegraphics[width=8cm]{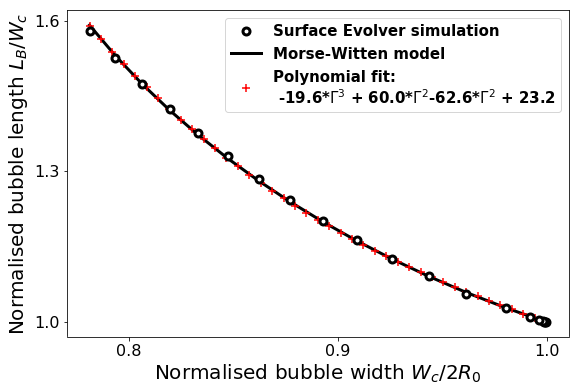}
\caption{Variation of the aspect ratio $ \lambda = \frac{L_B}{W_C}$ with the confinement ratio $\frac{W_C}{2R_0}$ for isolated bubbles in a capillary, in the absence of gravity.  Surface Evolver simulations and the prediction of Eq. (\ref{eq:aspectratio}) from Morse-Witten theory (full black line) are in excellent agreement. For practical purposes, we fit a third-order polynomial to the curve.}
\label{fig:IsolatedBubble}
\end{figure}

To check that the mesh distortion does not stall the convergence procedure in the case of small contact forces, we verify that, starting from the initial structure illustrated in Figure \ref{fgr:SE_Relax},  the algorithm yields as expected a null contact force in the case where the undeformed bubble diameter is equal to the distance between opposite walls and to the distance between neighboring bubbles centers, i.e. when $W_C/2R_0 = L_B/2R_0 = 1$. We also perform Surface Evolver simulations of isolated bubbles which are confined only by the capillary walls, but not by neighbouring bubbles, i.e. $W_C/2R_0 < 1$ and $L_B/2R_0 > 1$. For this purpose, the constraints on the faces $f_B$ are removed during the simulation. Fig. \ref{fig:IsolatedBubble} shows how in this case the bubble  aspect ratio $\lambda = \frac{L_B}{W_C}$ varies with the confinement ratio $\frac{W_C}{2R_0}$. In the investigated range of confinement ratios, these results are in full agreement with the analytical prediction Eq. (\ref{eq:aspectratio}) without free parameters, derived from Morse-Witten theory. For practical purposes (Section \ref{sec:Methods}),  it can be fitted by a polynomial. 
\begin{equation}
\lambda = \frac{L_B}{W_C}\approx -19.6 \left( \frac{W_C}{2R_0}\right)^3+60 \left( \frac{W_C}{2R_0}\right)^2-62.6 \left(\frac{W_C}{2R_0}\right)+23.2
\label{eq:IsolatedBubbles}
\end{equation}

\section{Experimental methods and procedures}
\label{sec:Methods}
\subsection{Materials and bubble generation}
For the experimental investigations we use Sodium Dodecyl Sulfate solutions at 7 g/L which corresponds to 2.9 times the critical micellar concetration (CMC). This concentration is high enough to ensure bubble stability and  to neglect depletion of surfactant by adsorption to the interfaces. Yet, as discussed in more detail in Appendix \ref{sec:depletionForces}, it is low enough to neglect attractive depletion forces between bubbles which arise at high micelle concentration. Non-negligible depletion forces would lead to finite contact angles in experiments which we neglected in the modeling. 

Solutions are freshly prepared in Millipore water every two days by stirring for 30 min. The surface tension of the solution is measured to be $\gamma = 0.034$ N/m at room temperature (20 \degree C) using a pendant drop device (TRACKER from TECLIS). 

\begin{figure}[ht!]
\centering
\includegraphics[width=8.5cm]{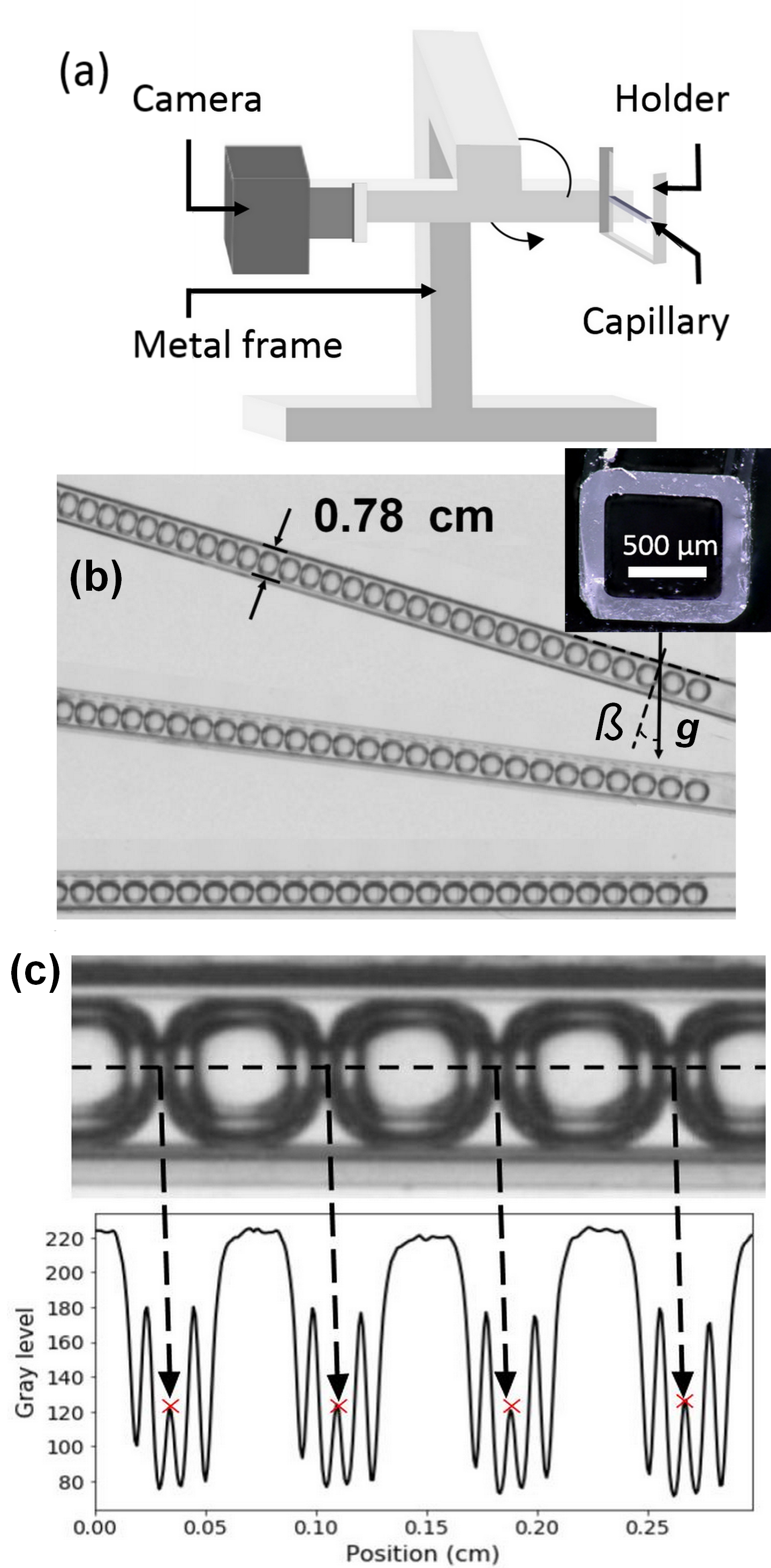}
\caption{(a) Experimental set-up: camera and capillary are held by the same metal arm which can rotate to tilt the capillary with respect to gravity. (b) Examples of photographs of bubble trains in capillaries ($W_C = 0.78$ mm) at three different tilt angles $\beta = 17$\degree. Insert: microscope image of the cross-section of the capillary. (c) Illustration of image treatment used to obtain the bubble length $L_B$. Top: profile view of bubbles trapped in capillary. Bottom: gray value profile with detection of bubble boundary (red cross).}
\label{fig:Experiment}
\end{figure}

Bubbles are produced by blowing air at constant pressure (Elveflow pressure controller PG1113, P = 11 mbar) into the SDS solution through needles (Nordson EFD) with circular cross-sections of different inner radii $R_C$ ($R_C = 150-330$ $\mu$m). For sufficiently small gas flow rates, the generated bubble radius $R_0$ is proportional to $R_C^{-3}$, which therefore serves to control the bubble size. The generated bubbles are trapped in glass capillaries with square cross-section of different internal widths $0.6$ mm $\leq W_C \leq 1$ mm (Vitrocom), whose dimensions are systematically verified using a Keyence numerical microscope (Insert of Fig. \ref{fig:Experiment}, Keyence VHX5000). The bubbles are trapped  manually by holding the end of the capillary above the point of bubble generation. Once the capillaries are filled with about 20 bubbles, they are sealed at either end using Blu-Tack adhesive paste. Each capillary can be used for about 3 h before gas exchange between the bubbles leads to measurable bubble-size variations.

\subsection{Experimental set-up et procedure}
A schematic of the overall experimental set-up is shown in Figure \ref{fig:Experiment}a. The square capillary is attached to a metal frame which also holds the digital camera (IDS UI-3580LE and TAMRON M118FM50 camera lens). It fixes the relative positions of the camera and the capillary. The entire frame can be rotated with respect to gravity. The capillary is imaged in front of a diffusive white screen with homogeneous lighting. The latter is placed 40 cm behind the capillary in order to benefit from optical effects which make the bubble boundaries appear dark black \cite{VanDerNet_Blondel_2006}. A thin wire with an attached weight suspended in the field of view of the camera is used to detect the vertical direction in the images. The precise angle $\beta$  between the normal of the capillary and the direction of gravity (Fig. \ref{fig:Experiment}b) is then obtained from image analysis. 

Every time the angle is varied, we wait until there is no more measurable change between two consecutive images taken at an interval of 5 min. This equilibration takes 15-30 minutes, depending on the inclination angle $\beta$. Figure \ref{fig:Experiment}b shows examples of images obtained at 3 different angles.\par
To obtain the length $L_B$ of each bubble along the bubble train, we use the image analysis program ImageJ \footnote{https://imagej.nih.gov/ij/} to measure the profile of gray values along the centre line of the capillary - as shown in Figure \ref{fig:Experiment}c. Due to optical effects, the contact zones between neighbouring bubbles appear as three bright spots, surrounded by dark areas. The actual border between two bubbles is the central bright spot\cite{VanDerNet_Blondel_2006}. A home-made Python algorithm detects the central spot, indicated by red crosses in Figure \ref{fig:Experiment}c.

The volume of the bubbles in the capillaries is determined by measuring their length $L_B$ at $\beta = 0$ degree. This length is then converted into the undeformed bubble radius $R_0$ using Eq. (\ref{eq:IsolatedBubbles}) obtained from fitting Surface Evolver simulations and theory of isolated bubbles in capillaries (see Section \ref{sec:evolver} and Figure \ref{fig:IsolatedBubble}). \\

\subsection{Force calculation}
The contact force $F_B(n)$ at the bottom of each bubble $n$ (counted from the bottom bubble with the first bubble being $n = 1$) is obtained by calculation of the buoyancy force exerted by the $n-1$ bubbles underneath, i.e.
\begin{equation}
F_B(n) = sin(\beta)g\rho(n-1)\frac{4}{3}\pi R_0^3,
\end{equation}
where $g$ is the gravitational acceleration and $\rho$ is the density of water. We calculate the overall force $F(n)$ exerted on the n-th bubble as the average of the force exerted on its bottom and top contact 
\begin{equation}
F(n)= \frac{1}{2} \left( F_B(n) + F_B(n+1) \right).
\end{equation}
If the pressure variations across one bubble are negligible, $F_B(n)=F(n)$. This is the assumption in the Morse-Witten model derived in Section \ref{sec:MW} and we shall show in Section \ref{sec:results} and APPENDIX \ref{sec:theorygravity} that this is indeed a good approximation for the range of experimental parameters investigated here. 
In the following, we present the normalised force per bubble, given as
\begin{equation}
f(n) = \frac{F(n)}{\gamma R_0}.
\end{equation}
In the experiments we have access only to the bubble-bubble forces, but not to the bubble-wall forces. Nevertheless, bubble-wall forces are consistently provided through the modeling. They are normalised in the same manner.

\section{Results and Discussion \label{sec:results}}

We first discuss the predictions of the two-body interaction model (Section\ref{sec:twobody}), of the model based on Morse-Witten theory (Section \ref{sec:MW}) and of the Surface Evolver simulations (Section \ref{sec:evolver}). We assume in each case that gravity-induced pressure gradients in the continuous phase are negligible on the scale of a bubble ($Bo \rightarrow 0$) - which will be shown later to be a reasonable approximation within the range of investigated experimental parameters. Since the Surface Evolver simulations rely on the numerical solution of the Laplace equation without additional approximations, we use them as a reference to check the analytical models. 

\begin{figure*}
\centering
  \includegraphics[width=\textwidth]{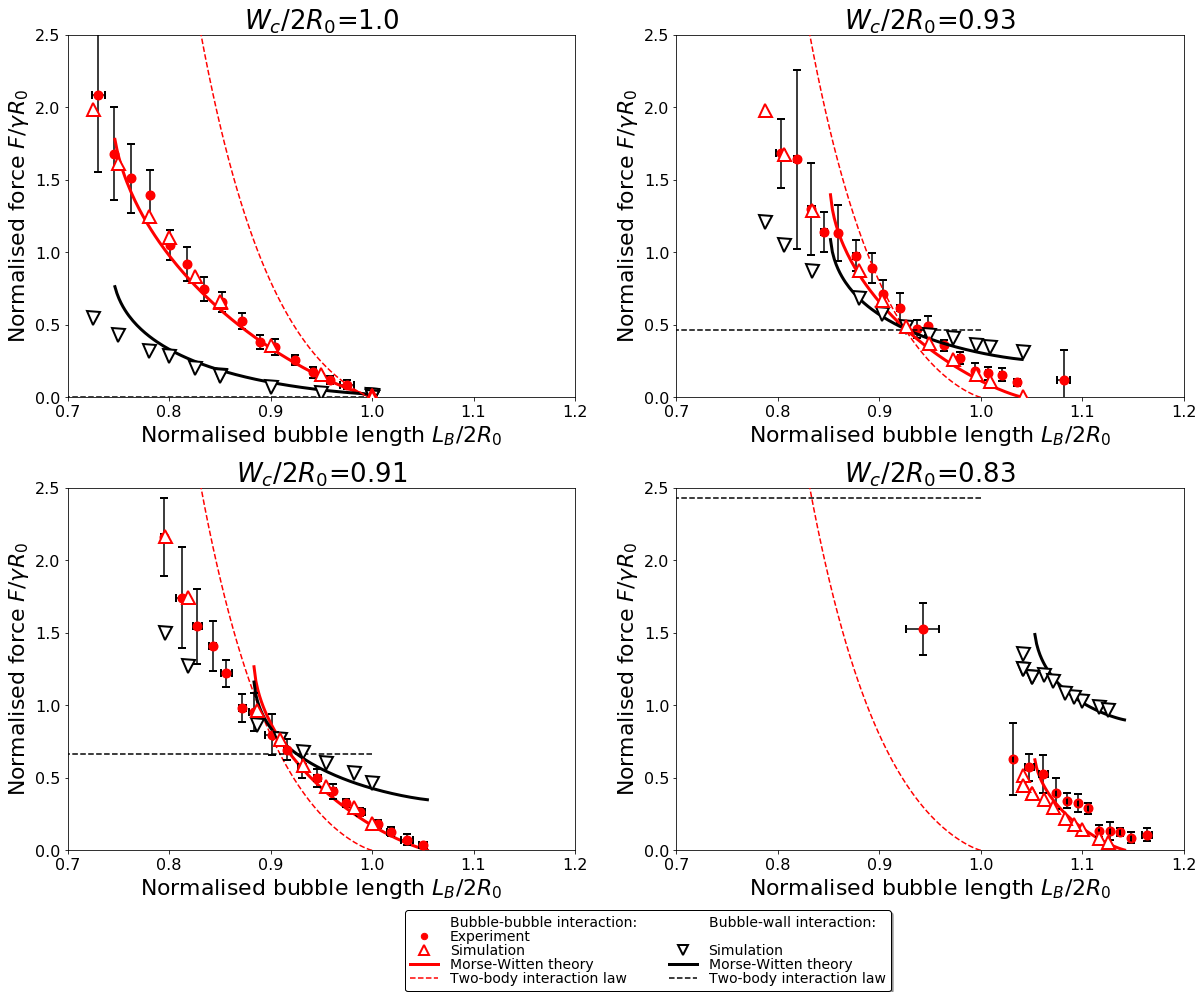}
 \caption{ Overview of experimental results, simulations and model predictions, for four different confinement ratios, as indicated. The normalised bubble-bubble and bubble-wall contact forces are plotted versus the normalised bubble length along the capillary axis. The experimental data are averages of data obtained for different inclination angles of the capillary $\beta$ ranging from 0 to 30 degrees and for different Bond numbers in the range of $0.044 < Bo < 0.087$.The 4th graph ($W_C/2R_0=0.83$) is the limit case for the hypothesis $f<1$ to be valid. }\label{fig:ExpResults}
\end{figure*}

Figure \ref{fig:ExpResults} shows how the normalised bubble-bubble (red) and bubble-wall forces (black) $F/\gamma R_0$ depend on the normalised bubble length $L_B/2R_0$ for confinement ratios in the range $0.83 \leq W_C/2R_0 \leq 1$. Morse-Witten theory is in good agreement with the Surface Evolver simulations all over the predicted range of validity ($F/\gamma R_0 < 1$). For a confinement ratio of $1$, for which the undeformed bubble fits exactly into the capillary, all contact forces go to zero when the bubbles stop touching their neighbours at $L_B/2R_0 =1$, as expected. As the bubble train is compressed along the axis of the capillary, bubble-bubble forces build up and, at the same time, bubble-wall forces appear since the bubbles expand laterally and push on the wall. This latter effect is ignored by the two-body interaction model which also over-predicts the bubble-bubble forces.  For confinement ratios smaller than 1, the undeformed bubble radius $R_0$ is too large for spherical bubbles to fit into the capillary. They therefore exert wall forces even if  the bubble-bubble deformation $x_B$  is zero. This effect is shown clearly by the Surface Evolver simulations and predicted quantitatively by the Morse-Witten theory - in contrast to the two-body interaction model. \par

For the confinement ratios 0.91 and 0.93, there is a specific value of the normalised bubble length $L_B/2R_0$ where the bubble-bubble and the bubble-wall forces coincide. This is the only case where the two-body interaction model provides the correct prediction. This is indeed expected, since the free parameters of the two-body model were fitted to Surface Evolver simulations of bubbles subjected to {\it isotropic} compression \cite{Lacasse1996}. For the smallest investigated confinement ratio of 0.83, even the smallest bubble-wall contact forces are already close to 1. Since this is the strongest confinement that can be handled by the Morse-Witten theory, one sees a rapid divergence between simulation and theory. Smaller confinement ratios are therefore not investigated in the following in order to focus on experiments which satisfy the approximations made in the theory.\par

\begin{figure*}[h!]
\centering
\includegraphics[width=\linewidth]{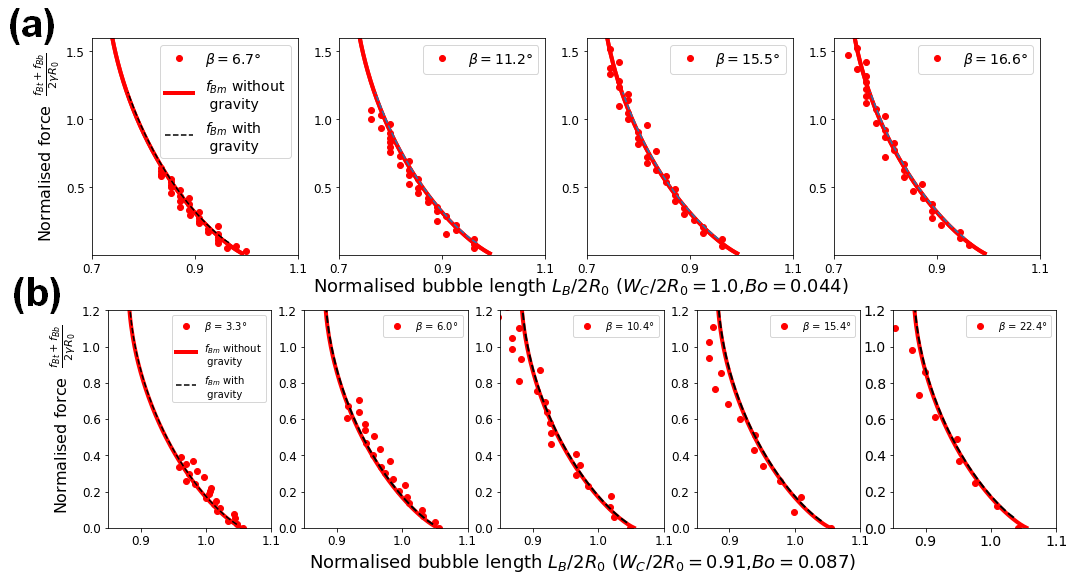}

\caption{Examples of measurements of the normalised bubble-bubble force $F/\gamma R_0$ vs normalised bubble length $L_B/2R_0$, obtained for different capillary angles $\beta$. (a) shows data obtained for the lowest and (b) for the highest Bond number encountered in our experiments (a: $Bo$ = 0.044,$W_c=0.775$ mm, $\frac{W_c}{2R_0}=1.0$, b: $Bo$ = 0.087,$W_c=1$ mm, $\frac{W_c}{2R_0}=0.91$). The full red line represents the Morse-Witten theory given in Eq. (\ref{eq:aspectratio}), neglecting pressure gradients in the liquid. The dashed line show predictions derived in APPENDIX A where pressure gradients are taken explicitly into account. Manifestly, gravity has no significant impact here.}
\label{fig:forcevsheight}
\end{figure*}

Let us now turn to the experiments. Since Bond numbers $Bo=\rho g {R_0}^2/\gamma$ in our experiments are in the range of $0.044 < Bo < 0.087$ one needs to analyse the potential impact of gravity-induced pressure variations over the length scale of an individual bubble which are neglected in the models presented in Section \ref{sec:TheoSim}. The four wall contacts are then no longer equivalent by symmetry, and the  contact forces exrted on upper neighbor or capillary wall are different from the bottom ones. Both of these effects depend on the angle of inclination of the capillary and on the Bond number Bo. To asses the importance of this effect in the experimental data, we explicitly show in Fig. \ref{fig:forcevsheight} data obtained for different tilt angles for lowest and highest Bond number encountered in our experiments (Bo = 0.044 in Fig. \ref{fig:forcevsheight}a and Bo = 0.087 in Fig. \ref{fig:forcevsheight}b). One observes that within the experimental scatter, the data for all angles is well described by the Morse-Witten theory given in Eq. (\ref{eq:aspectratio}) without any systematic deviation. This observation confirms the predictions made by the extension of Morse-Witten theory which includes pressure gradients over the scale of one bubble, which we develop and discuss in APPENDIX \ref{sec:theorygravity}. The predictions are plotted in Figure \ref{fig:forcevsheight} as dashed line for comparison. One can see that this effect is negligible over the range of Bond numbers investigated.  APPENDIX \ref{sec:theorygravity} shows that pressure gradients should be taken into account in this experiment for $Bo > 0.15$. Since all our Bond numbers are smaller than this value, we consider pressure gradients on the bubble scale as negligible in the remaining discussion and we compare systematically with the the models and simulations developed without gravity in Section \ref{sec:TheoSim}. \par

With pressure gradients along one bubble being negligible, we can average all data obtained for the same confinement ratio $W_C/2R_0$, but different angles $\beta$ and capillary widths $W_C$. For a given bubble length. Fig. \ref{fig:ExpResults} therefore presents experimental results for four different confinement ratios, which are averaged over at least 10 different data sets on 10-20 bubbles each for different $\beta$ and $W_C$. Figure \ref{fig:ExpResults} shows that the obtained experimental results agree very well with  Morse-Witten theory and the simulations for dimensionless contact forces up to unity, this limit being set by the approximations made in the model (Section \ref{sec:MW}). For larger forces, we see good agreement between the experiments and the Surface Evolver simulations. As pointed out previously, the two-body interaction model fails to predict the experimental observations, except in the special case where the contact forces are all equal.

\section{Conclusion and outlook}

Our experiments, analytical calculations and Surface Evolver simulations consistently show that the interactions of bubbles are non-pairwise: the mechanical response at a given contact is a function of the confinement of the bubble at all its other contacts, either by walls or by neighboring bubbles. Our results also demonstrate that this effect is captured quantitatively by an analytical interaction model derived from the linearised Laplace equation, within the framework of  Morse-Witten theory. The range of validity of this theory is limited to contact forces smaller than $\gamma R_0$, as expected. We also show that effective two-body  models do not capture the non-pairwise interactions of real bubbles.\par

 The Morse-Witten theory of bubble contacts was initially developed for foams which are topologically ordered and only very weakly polydisperse \cite{Hohler2017}. Recently, the theory has  been extended to predict the contact forces in disordered polydisperse wet foams \cite{Hohler2018}. The validation of this generalised interaction model is an important perspective for further experimental work. 

Our study was performed for bubbles small enough that the deformation due to the pressure variation in the water induced by gravity was negligible compared to that induced by contact forces (Bo < 1). However, in Appendix \ref{sec:theorygravity} we show how Morse-Witten theory can be used to include such pressure variations in an iterative scheme. This will be of interest for other experiments.\par

The experiments and simulations reported here provide a benchmark test for models aiming to represent bubbles and droplets as soft particles, and they have highlighted the importance of the non-pairwise coupling of the contacts. A natural extension of this work concerns quasi-2D foams or emulsions where bubbles or droplet monolayers, surrounded by liquid, are confined between the parallel walls of a Hele-Shaw cell. Provided that the confinement by the cell wall is not too strong, Morse-Witten theory provides a basis for predicting the structure, osmotic pressure and static mechanic response of such systems, depending on the confinement ratio. Unfortunately, this criterion was not satisfied by previous experiments\cite{desmond2013} in this configuration, from which an empirical interaction law was derived. A simple example of a quasi-2D system can be directly derived from the work reported in the present paper: Let us consider the individual bubble that we have modeled as a repeat unit ("unit cell", in the terminology of crystallography). Stacking it up in a plane provides the structure of an ordered (quadratic) quasi-2D foam. Stacking it up in 3 dimensions yields a cubic bubble crystal, previously studied using Morse-Witten theory in the case of infinitesimal deformations by Buzza {\it et al}\cite{Buzza1994}. However, these cubic structures are known to be unstable to shear, and much further work is needed to apply Morse Witten theory to 3D ordered or disordered foam and emulsion structures, as discussed in detail in \cite{Hohler2018}.

Our results validate Morse Witten theory which provides a basis for future simulation models of wet foams or emulsions. The interaction law applies to arbitrary coordination numbers and contact positions that may occur in a disordered weakly polydisperse system with zero wetting angle, with or without confining walls. A simulation model on this basis, in the spirit of molecular dynamics, would represent an order of magnitude gain in computational efficiency, compared to Surface Evolver simulations, for the same level of accuracy, thus allowing a quantitative comparison with experiments. Setting up such a simulation model involves technically challenging issues, but the expected benefit is substantial: It could provide insight about the impact of non-pairwise interactions on jamming,  yielding and flow of real foams and emulsions. The coupling among the contacts can be expected to have an impact on the bubble contact statistics which are a key issue for the jamming transition \cite{VanHecke2010, Liu2010, Charbonneau2014}. Simulations based on the Morse-Witten interaction model could also be used to model the flow of strongly confined bubbles and drops, relevant in many microfluidic applications.\par
For future work it will be important to model interactions for more strongly deformed bubbles and for the case of finite contact angles.

\section{Acknowledgements}
The authors acknowledge stimulating discussions with Denis Weaire, Stefan Hutzler and Emmanuelle Rio. We thank Robin Bollache and Leandro Jacomine for help with the experiments. Research at the INSP was funded by the European Space Agency (Soft Matter Dynamics MAP grant AO-09-943). The ICS acknowledges funding from the European Research Council (ERC) under the European Union's Seventh Framework Program (FP7/2007-2013) in form of an ERC Starting Grant, agreement 307280-POMCAPS. This work is also published within the IdEx Unistra framework and has benefited from funding from the state, managed by the French National Research Agency as part of the 'Investments for the future' program.



\bibliographystyle{rsc}
\bibliography{rsc
}

\newpage
\appendix



\section{Model of a bubble train in a tube in the presence of gravity}
\label{sec:theorygravity}

\begin{figure*}[ht!]
\centering
\includegraphics[width=18cm,keepaspectratio]{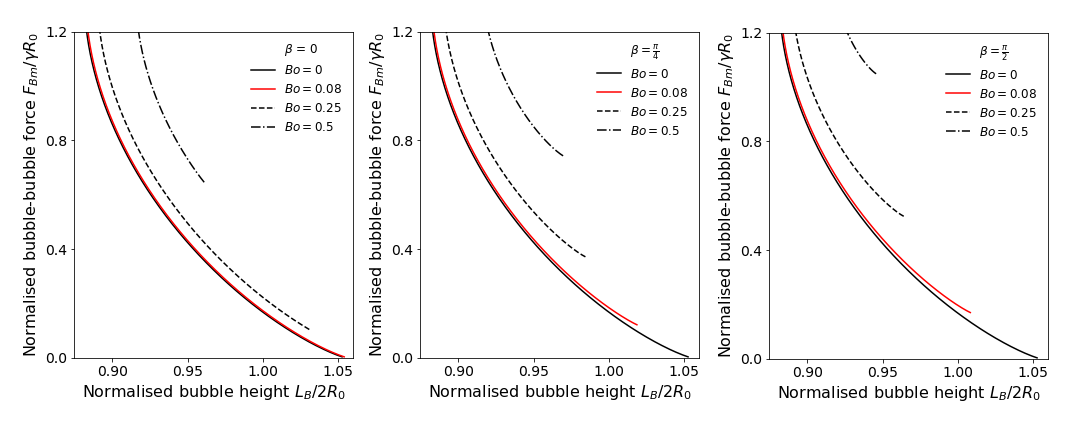}
\caption{Inter-bubble contact forces ($f_{Bm}$ in the equations) versus dimensionless bubble length along the axis of the tube predicted by Morse-Witten theory.The confinement ratio is 0.91.  The behavior in the absence of pressure gradients, analyzed in section \ref{sec:MW} is given by the black line which is identical on the three plots. The effect of a gravity induced pressure gradient in the liquid depends on the Bond number $Bo$ and on the inclination of the bubble train with respect to gravity $\beta $, as indicated on the figures and discussed in Section \ref{sec:MW}. \label{fig:Theorypicture}}
\label{fgr:example2col}
\end{figure*}
\begin{figure*}[h!]
\centering
\includegraphics[width=12cm]{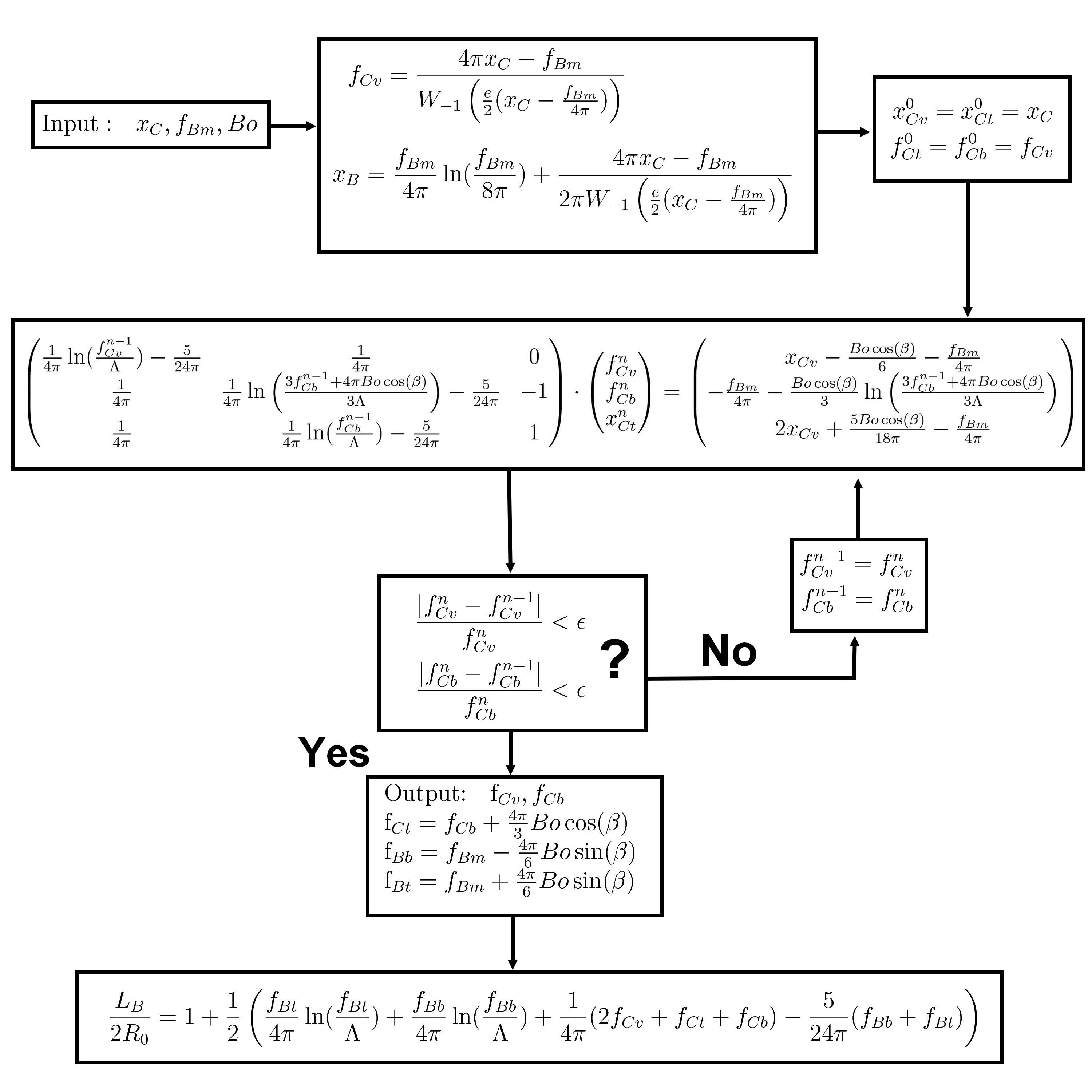}
\caption{Schematic of the autocoherent algorithm used to solve Eqs. (\ref{eq:nonlinear2}). $\epsilon$ is an arbitraty threshold which is tuned according to the desired precision.}
\centering
\label{schematicautocoherent}
\end{figure*}

We now generalize our analysis to the case where a pressure gradient is present in the liquid, due to gravity. We study the case where the tube is inclined by an angle $\pi/2-\beta $ with respect to the direction of gravity (Fig \ref{fig:Experiment} ), in such a way that two of the capillary sidewalls always remain parallel to this direction. We introduce a notation that distinguishes forces and displacements at the top inter-bubble contact ($f_{Bt}, x_{Bt}$), and the bottom inter-bubble contact ($f_{Bb}, x_{Bb}$), on the top and the bottom tilted capillary walls  ($f_{Ct}, x_{Ct}$ and $f_{Cb}, x_{Cb}$) and on the two capillary walls that remain vertical($f_{Cv}, x_{Cv} $). These two latter contacts are equivalent by symmetry. In view of our experiments reported in the following, the aim of our calculation is to predict the bubble length along the axis of the capillary 
\begin{equation}
\label{eq:LB}
\frac{L_B}{2 R_0} = 1+\frac{x_{Bt}+x_{Bb}}{2}
\end{equation}
for given forces $f_{Bt}, f_{Bb}$ and confinement ratio 
\begin{equation}
\label{eq:constraint1}
\frac{W_C}{2R_0}=1+x_{Cv}=1+(x_{Ct}+x_{Cb})/2
\end{equation}
In mechanical equilibrium the total dimensionless buoyancy force acting on the bubble must be equilibrated  by its six contact forces, depending on the angle $\beta $ and the Bond number $Bo$. This leads to the relations
\begin{align}
 f_{Ct}- f_{Cb}=\frac{4\pi}{3} Bo\cos \beta \label{eq:constraint2}\\
  f_{Bt}- f_{Bb}=\frac{4\pi}{3} Bo \sin \beta.\label{eq:constraint3}
\end{align}
 The interaction law \ref{eq:interactionlaw} yields the following five equations, relating the contact forces to the contact displacements.
\begin{align}
x_{Bt}=  \frac{1}{4\pi} \ln\left(\frac{f_{Bt}}{\Lambda}\right) f_{Bt} -G(\frac{\pi}{2})(  2f_{Cv}+f_{Ct} +f_{Cb})  - G\left(\pi\right) f_{Bb} \label{eq:Bt}\\
x_{Bb}=  \frac{1}{4\pi} \ln\left(\frac{f_{Bb}}{\Lambda}\right)  f_{Bb} -   G(\frac{\pi}{2})(  2f_{Cv}+f_{Ct} +f_{Cb})- G\left(\pi\right) f_{Bt}\label{eq:Bb} \\
x_{Cv}= \frac{1}{4\pi} \ln\left(\frac{f_{Cv}}{\Lambda}\right)  f_{Cv} -    G(\frac{\pi}{2})  (f_{Ct}+f_{Cb}+f_{Bt}+f_{Bb})-   G(\pi)  f_{Cv} \label{eq:xCv} \\
x_{Ct}= \frac{1}{4\pi} \ln\left(\frac{f_{Ct}}{\Lambda}\right)  f_{Ct} -    G(\frac{\pi}{2})  (2f_{Cv}+f_{Bt}+f_{Bb})-   G(\pi)  f_{Cb}\label{eq:xCT}\\
x_{Cb}= \frac{1}{4\pi} \ln\left(\frac{f_{Cb}}{\Lambda}\right)  f_{Cb} -    G(\frac{\pi}{2})  (2f_{Cv}+f_{Bt}+f_{Bb})-   G(\pi)  f_{Ct}\label{eq:xCb}
\end{align}
We have introduced the constant $\Lambda=8\pi e^{-5/6}$ to write these equations in a more concise way. In view of Eq. (\ref{eq:LB}), adding Eqs. (\ref{eq:Bt}) and (\ref{eq:Bb})  provides an expression for the bubble contact displacements along the tube axis as a function of the sum of the wall and inter-bubble contact forces.
\begin{multline}
x_{Bt}+x_{Bb}=2\left(\frac{L_B}{2R_0}-1\right)\\
=\frac{1}{4\pi} \ln\left(\frac{f_{Bt}}{\Lambda}\right) f_{Bt}+ \frac{1}{4\pi} \ln\left(\frac{f_{Bb}}{\Lambda}\right)  f_{Bb} \\ + \frac{1}{4\pi}(  2f_{Cv}+f_{Ct} +f_{Cb}) -  \frac{5}{24\pi}  (f_{Bb}+f_{Bt})
\end{multline}
The only unknowns on the right hand side of this expression, $ 2f_{Cv}+f_{Ct}$, must be deduced from Eqs. (\ref{eq:xCv}) - (\ref{eq:xCb}). In these three equations, there are three unknown quantities $f_{Cv},f_{Cb},x_{Ct}$, all the other forces and displacements appearing in Eqs. (\ref{eq:xCv}) - (\ref{eq:xCb}) can be expressed in terms of $f_{Cv},f_{Cb},x_{Ct}$ using Eqs. (\ref{eq:constraint1}) - (\ref{eq:constraint3}). This yields the set of equations
\begin{multline}
\label{eq:nonlinear2}
x_{Cv}= \frac{1}{4\pi} \ln\left(\frac{f_{Cv}}{\Lambda}\right)  f_{Cv}     +\frac{Bo\cos \beta}{6}+\frac{f_{Cb}+f_{Bm}}{4\pi}-  \frac{5}{24 \pi}  f_{Cv} \\
x_{Ct}=  \ln\left(\frac{f_{Cb}+\frac{4\pi}{3} Bo\cos \beta}{\Lambda}\right)  \left(\frac{Bo\cos \beta}{3} +\frac{f_{Cb}}{4\pi}\right) +\frac{f_{Cv}+f_{Bm}}{4 \pi} -  \frac{5}{24 \pi}  f_{Cb}\\
2x_{Cv}- x_{Ct}= \frac{1}{4\pi} \ln\left(\frac{f_{Cb}}{\Lambda}\right)  f_{Cb} +\frac{f_{Cv}+f_{Bm}}{4 \pi}  -   \frac{5}{24 \pi}  f_{Cb}-
 \frac{5}{18 \pi}Bo\cos \beta
\end{multline}
To make these expressions more concise we have introduced the mean bubble-bubble force $f_{Bm}=(f_{Bt}+f_{Bb})/2$.
We solve this nonlinear set of equations numerically, as in previous work \cite{Hohler2017}. We do this iteratively, starting from the rough approximation that all wall contact forces $f_{Ct},f_{Cb},f_{Cv}$ are equal to the prediction of Eq. (\ref{eq:XCf2}), strictly valid in the limit of small Bond numbers.  We substitute in the logarithmic terms of Eq. (\ref{eq:nonlinear2}) these estimates of $f_{Cb}$  and $f_{Cv}$. This reduces these three non-linear equations to a set of linear  equations which we solve for $f_{Cb}$, $x_{Ct}$  and $f_{Cv}$ by standard methods. We thus obtain an improved estimate for  $f_{Cb}$, $x_{Ct}$  and $f_{Cv}$ and use it to update the logarithmic terms in Eq. (\ref{eq:nonlinear2}). This procedure is iterated until convergence is achieved. For the iteration, the set of three equations Eq. (\ref{eq:nonlinear2}) can be reduced to two equations which are more complex. Details of the numerical algorihm are given in Figure \ref{schematicautocoherent}.\par
Figure \ref{fig:Theorypicture} illustrates the effects of gravity and pressure gradients on a bubble train confined in a tube predicted by this calculation, based on Morse-Witten theory. The black line in all three plots (often hidden by almost identical lines with other colors) shows the behaviour in the case where the pressure is homogeneous in the liquid ($Bo =0$),the case studied in section \ref{sec:MW}. In the presence of gravity ($Bo > 0$), the bubbles are squeezed all over their gas-liquid interfaces by buoyancy forces depending on position. They act in addition to the contact forces exerted by the neighbouring bubbles.
If the capillary is held horizontally ($alpha =0$), the bubbles are squeezed against the top confining wall, leading to an increase of the bubble length $L_B$ measured along the axis of the capillary. However,\ref{fig:Theorypicture} shows that this effect is negligible for Bond numbers up to 0.075 and a typical confinement ratio 0.91 relevant for our experiments. If the capillary is held vertically ($\beta = \pi/2$), the length $L_B$ decreases with increasing Bond numbers, for inter-bubble contact forces down to zero. This effect, due to the pressure gradient in the liquid, is similar to the deformation of a bubble surrounded by liquid, buoyed against an immersed  horizontal plate, studied in Morse and Witten's original work. In our case, this deformation is modified by the presence of confining walls. The inter-bubble force, defined here as the average of the top and the bottom contact force, cannot be arbitrarily small for ($\beta = \pi/2$), since even if the force at the bottom is zero because we consider the lower end of a bubble train,  a force at the top contact is always required to maintain static equilibrium. This effect is clearly seen on Fig \ref{fig:Theorypicture}. It also shows that for contact forces larger than this minimal value, the results are very close to those obtained in the absence of a pressure gradient, for the  range of Bond numbers relevant for our experiments.   The results for ($\beta = \pi/4$), also shown on on Fig. \ref{fig:Theorypicture}, are intermediate between the results for ($\beta = 0$) and ($\beta = \pi/2$). \par 
To summarize, our calculations derived from Morse Witten theory show that the effects of pressure gradients do not modify the relation between inter-bubble forces and bubble length $L_B$ in the range of Bond numbers relevant for our experiments, for all inclination angles $\beta $.

\section{Depletion interactions between bubbles}
\label{sec:depletionForces}
The attractive depletion force $F(h)$ between two spherical bubbles of radius $R$ whose surfaces are separated by a distance $h$ in a liquid containing spherical micelles of radius $r$ can be calculated using the Derjaguin approximation \cite{Derjaguin_Kolloid_1934,Oettel_PRE_2004,Kralchevsky_COCIS_2015}
\begin{equation}
F_D(h) = \pi(R+r)\left[P_{mic}(h-r)-2\sigma_{\inf}\right],
\end{equation}
where $P_{mic}$ is the pressure exerted by the micelles on the bubble surface and $\sigma_{inf}$ is the excess surface free energy of a hard-sphere fluid. Both can be expressed in terms of the micelle volume fraction $\phi$
\begin{align}
P_{mic} &= \frac{kT}{r^3}\frac{6}{\pi}\phi\frac{1+\phi+\phi^2-\phi^3}{(1-\phi)^3}\\
\sigma_{inf}&=-\frac{kT}{r^2}\frac{9}{2\pi}\phi^2\frac{1+\phi}{(1-\phi)^3}.
\end{align}
Here $kT$ is the thermal energy.
Using these expressions we can estimate the order of magnitude of depletion forces acting in our system. The critical micellar concentration of SDS is 8 mM, meaning that the SDS solution contains 3*8 mM of surfactant molecules. The typical aggregation number of SDS is of the order of 50, while the characteristic radius of SDS micelles is about $r=$ 0.2 nm \cite{Danov_ACIS_2011, Pisarcik_OpenChem_2015}. Making the approximation that all surfactant molecules are contained in micelles, we can estimate the volume fraction of the micelles as $\phi \approx 10^{-3}$. Taking $T = 293$ K (room temperature) and $R \approx 0.5$ mm, one finds that the largest possible normalised depletion forces $F_D/\gamma/R_0$ (which arise for $h=0$) are of the order of $10ç{-3}$. This is two orders of magnitude smaller than the characteristic forces measured in our experiments and hence negligible with respect to the experimental errors.

The above calculations use hard-sphere approximations to estimate the depletion forces acting between objects in micellar SDS solutions, as has been frequently done in the past \cite{Bibette_PRL_1990,Jorjadze_PNAS_2011}. However, care needs to be taken in the case of SDS since it is a negatively charged surfactant. The electrostatic repulsion between the charged surfaces in the absence of salt ensures that the two bubble surfaces are kept at distances of at least 20 nm for the concentrations used here \cite{Danov_ACIS_2011,Kralchevsky_COCIS_2015}. Since the micelles are much smaller than this distance, one cannot truely speak about an excluded volume effect. Instead, since the micelles are also charged, one needs to take into account the long-range electrostatic interactions and counterion effects. This can lead to attractive interactions which increase with increasing surfactant concentration. However, they are negligibly small for the surfactant concentration used here (3 CMC) \cite{Danov_ACIS_2011,Kralchevsky_COCIS_2015}.

\end{document}